\documentclass[a4paper,reqno]{amsart}
\usepackage[all]{xy}           %For commutative diagrams
\usepackage{amssymb}           %For double arrows
\usepackage{hyperref}
\usepackage{eucal}
\usepackage{graphicx}
\usepackage{epsfig}
\usepackage[usenames,dvipsnames]{color}
\usepackage{color}
\numberwithin{equation}{section}

\newtheorem{definition}{Definition}[section]

\newtheorem{theorem}[definition]{Theorem}

\newtheorem{remarkth}[definition]{Remark}

\newenvironment{remark}{\begin{remarkth}\upshape}{\hfill$\diamond$\end{remarkth}}
\renewcommand{\emph}[1]{{\bfseries\itshape{#1}}}
\numberwithin{figure}{section}

\newcommand{\R}{\mathbb{R}}      %Numeros reales
\newcount\ancho \newcount\anchom \newcount\anchoa
\newcount\anchob \newcount\altura

\newcommand{\ltilde}[3][0]{\altura=0 \advance\altura by #1
           \ancho=#2 \anchom=\ancho \divide\anchom by 2
           \anchoa=\ancho \divide\anchoa by 4
           \anchob=\anchom \advance\anchob by \anchoa
           \kern-3pt \begin{array}[b]{c}
           \begin{picture}(1,1)(\anchom,-\altura)
        \qbezier(0,2)(\anchoa,5)(\anchom,2)
        \qbezier(\anchom,2)(\anchob,-1)(\ancho,4)
        \qbezier(0,2)(\anchoa,4.5)(\anchom,1.8)
        \qbezier(\anchom,1.8)(\anchob,-1.5)(\ancho,4)
       \end{picture} \\[-4pt]{#3}
                       \end{array} \kern-4pt    }

\newcommand{\lhat}[3][0]{\altura=0 \advance\altura by #1
           \ancho=#2 \anchom=\ancho \divide\anchom by 2
           \anchoa=\ancho \divide\anchoa by 4
           \anchob=\anchom \advance\anchob by \anchoa
           \kern-3pt \begin{array}[b]{c}
           \begin{picture}(1,1)(\anchom,-\altura)
        \qbezier(0,2)(\anchoa,4)(\anchom,6)
        \qbezier(\anchom,6)(\anchob,4)(\ancho,2)
        \qbezier(0,2)(\anchoa,3.8)(\anchom,5.6)
        \qbezier(\anchom,5.6)(\anchob,3.8)(\ancho,2)
       \end{picture} \\[-4pt] {#3}
                       \end{array} \kern-4pt    }

\newcommand{\I}{I\mkern-7muI}

\makeatletter
\newcommand\prol{\@ifstar{\@proldf}{\@prolpf}}  %% if * dual else primal
\def\@prolpf{\@ifnextchar[{\@prolpf@wrt}{\@prolpf@}}
\def\@prolpf@wrt[#1]#2{\@ifnextchar[{\@prolpf@wrt@at{#1}{#2}}{\@prolpf@wrt@{#1}{#2}}}
\def\@prolpf@wrt@at#1#2[#3]{\prolsymbol^{#1}_{#3}#2}
\def\@prolpf@wrt@#1#2{\prolsymbol^{#1}#2}
\def\@prolpf@#1{\@ifnextchar[{\@prolpf@at{#1}}{\@prolpf@@{#1}}}
\def\@prolpf@at#1[#2]{\prolsymbol_{#2}#1}
\def\@prolpf@@#1{\prolsymbol#1}
\def\@proldf{\@ifnextchar[{\@proldf@wrt}{\@proldf@}}
\def\@proldf@wrt[#1]#2{\@ifnextchar[{\@proldf@wrt@at{#1}{#2}}{\@proldf@wrt@{#1}{#2}}}
\def\@proldf@wrt@at#1#2[#3]{\prolsymbol^{*#1}_{#3}#2}
\def\@proldf@wrt@#1#2{\prolsymbol^{*#1}#2}
\def\@proldf@#1{\@ifnextchar[{\@proldf@at{#1}}{\@proldf@@{#1}}}
\def\@proldf@at#1[#2]{\prolsymbol^*_{#2}#1}
\def\@proldf@@#1{\prolsymbol^*#1}
\def\prolsymbol{\mathcal{T}}
\makeatother

\begin{document}
{\Large

\title{Non-existence of an invariant measure for a homogeneous ellipsoid rolling on the plane}

%\author[Yuri N.\ Fedorov]{Yuri N.\ Fedorov}
%\address{Yuri N.\ Fedorov:
%Department de Matematica Aplicada I \\
%Universitat Politecnica de Catalunya, Barcelona, E-08028 Spain}
%\email{Yuri.Fedorov@upc.edu}

\author[L. C. \ Garc\'{\i}a-Naranjo]{Luis C.\ Garc\'{\i}a-Naranjo}
\address{L. C.\ Garc\'{\i}a-Naranjo:
Departamento de Matem\'aticas y Mec\'anica \\
IIMAS-UNAM \\
Apdo Postal 20-726,  Mexico City,  01000, Mexico}
\email{luis@mym.iimas.unam.mx}

\author[J.\ C.\ Marrero]{Juan C.\ Marrero}
\address{Juan C.\ Marrero:
ULL-CSIC Geometr\'{\i}a Diferencial y Mec\'anica Geom\'etrica\\
Departamento de Matem\'atica Fundamental, Facultad de
Ma\-te\-m\'a\-ti\-cas, Universidad de La Laguna, La Laguna,
Tenerife, Canary Islands, Spain} \email{jcmarrer@ull.es}

\thanks{This work has been partially supported by MEC (Spain)
Grants MTM2009-13383, MTM2011-15725-E, MTM2012-34478 and the project of the Canary Government ProdID20100210. \\
LGN acknowledges the hospitality at the  Departamento de Matem\'atica Fundamental,
at Universidad de la Laguna, for his recent stay there. 
%JCM acknowledges the hospitality and support
%of the Department de Matem\'atica Aplicada I (UPC, Barcelona, Spain) and the Section de Math\'ematiques
%of EPFL (Lausanne, Switzerland) in several visits.
}

\keywords{nonholonomic mechanical systems, 
invariant volume forms, symmetries, reduction}

\subjclass[2010]{37C40,37J60,70F25,70G45,70G65}

\begin{abstract}
It is known that the reduced equations   for an axially symmetric homogeneous ellipsoid that rolls 
without slipping on the plane possess a smooth invariant measure. We show that such an invariant measure does not exist in the case when
all of the semi-axes of the ellipsoid have different length.
\end{abstract}

\maketitle

\tableofcontents

\section{Introduction}

The existence of a smooth invariant measure is a very important property for a system of 
autonomous ordinary differential equations. To date, there are a number of
research publications, e.g. \cite{ BM,  BME, BM2,  BBM, BMB, CaCoLeMa,Jo,Ko,Veselova,ZeBo} and others, that
analyze the existence of such an invariant for different mechanical systems with symmetry
that are subjected to nonholonomic constraints.

In a recent paper \cite{FeGaMa} we presented a general method and an algorithm  to examine the existence
of smooth invariant measure for the reduced equations of purely kinetic nonholonomic mechanical systems with symmetry. In this note
we apply the algorithm to show that the reduced equations of motion of a tri-axial 
homogeneous ellipsoid that rolls without slipping on the plane in the absence of gravity do not possess a smooth invariant measure. 

A related publication to this work is \cite{BME}. Here  the authors analyze the existence of an invariant measure  for a family of inhomogeneous ellipsoids rolling without slipping on the plane in the presence of gravity. Their method relies on the linear analysis of 
certain relative equilibria (vertical rotations) that only exist for special distributions of mass on the
ellipsoid (that do not contain the homogeneous distribution if the ellipsoid is tri-axial). We also mention that the rolling of an ellipsoid on the plane, with an additional
no-spin (rubber) constraint was recently considered in \cite{BMB}.

The configuration space for our system is $Q=SO(3)\times \R^2$. The $SO(3)$ part indicates the orientation
of the ellipsoid while the $\R^2$ part gives the $(x,y)$ coordinates of the center of the ellipsoid on
the plane where the rolling takes place. The symmetry group is the euclidean Lie group $SE(2)$ that corresponds to the
isotropy and homogeneity of the rolling plane.

\section{Algorithm to investigate the existence of invariant measures}
\label{S:algorithm}

We briefly recall the part of the algorithm presented in \cite{FeGaMa} to determine if the reduced equations of a nonholonomic system with symmetry
possess an invariant measure that is relevant for our system. It applies to  systems that satisfy the following conditions:
\begin{enumerate}

\item[{\bf C1.}] The first de Rham cohomology group of the shape space $\widehat Q=Q/G$ is trivial.

\item[{\bf C2.}] 
 There exists
an open dense set $\widehat U\subset \widehat Q$ with a global chart.
\end{enumerate}

In our example, the space
$\widehat Q=(SO(3)\times \R^2)/SE(2) \cong S^2$, that  satisfies both conditions {\bf C1} and  {\bf C2} (it is sufficient to take spherical coordinates on
$S^2$).

Let  $D\subset TQ$ be the non-integrable distribution on the configuration space $Q$ defined by the nonholonomic constraints. We denote by $p:Q\to \widehat Q$ the orbit projection and by $\mathcal{V}p\subset TQ$ the vertical
subbundle of $p$. That is $\mathcal{V}p(q)=T_q\mbox{Orb}_G(q)$ for all $q\in Q$, where $\mbox{Orb}_G(q)
\subset Q$
is the group orbit through $q$. We shall see that in our example the intersection
\begin{equation*}
\mathcal{V}^Dp(q):=D(q)\cap \mathcal{V}p(q)
\end{equation*}
has constant rank one, and that the {\em dimension assumption} (see \cite{BKMM}) $T_qQ=D(q)+ \mathcal{V}p(q)$ holds for all $q\in Q$. Let $\mathcal{G}$ denote the $G$-invariant Riemannian metric in $Q$ defined by
the kinetic energy of the system and define $\mathcal{H}:=(\mathcal{V}^Dp)^\perp\cap D$ where the
orthogonal complement  is taken with respect to $\mathcal{G}$. We note that 
$\mathcal{H}$ is the horizontal space of the nonholonomic connection defined in  \cite{BKMM}.

 In this case, the algorithm presented in \cite{FeGaMa} indicates the steps that are described below. In such a description, the latin subindices 
 $a, b, e$ run over the range of the vertical space $ \mathcal{V}^Dp$, the greek subindices $\alpha, \beta, \gamma$ run over the
 range of the horizontal space $\mathcal{H}$, and the capital latin subindices $I,J,K$ run over the joint range of $a, b, e$ and $\alpha, \beta, \gamma$.

\begin{enumerate}
\item[{\bf Step 1.}] Find a basis  $\{W_I\}=\{Z_a, Y_\alpha \}$ of $G$-invariant  vector fields  of $D$ in such a way
that $\{Z_a \}$ is a basis of sections of  $\mathcal{V}^Dp$ and $\{ Y_\alpha \}$ is a basis of sections of $\mathcal{H}$. In other words,
the vector fields $\{Z_a \}$ are $p$-vertical and we have $\mathcal{G}(Z_a,Y_\alpha)=0$ for all $a, \alpha$.

\item[{\bf Step 2.}] Compute the structure coefficients $C_{aI}^J$ defined by
\begin{equation*}
\mathcal P \left ( [ Z_a , Z_b ] \right )= C_{a b}^dZ_d + C_{a b}^\alpha Y_\alpha , \qquad 
\mathcal P\left (  [Z_a , Y_\alpha] \right )= C_{a \alpha}^bZ_b + C_{a \alpha}^\beta Y_\beta,
\end{equation*}
where $\mathcal P$ is the $\mathcal{G}$-orthogonal projection onto $D$ (that is $TQ=D\oplus D^\perp$ and $\mathcal P:TQ\to D$ is the orthogonal
projector) and $[\cdot, \cdot]$ is the standard commutator of vector
fields.
Notice that by $G$-invariance of the basis  $\{Z_a, Y_\alpha \}$ and the metric $\mathcal G$,
the structure coefficients $C_{aI}^J$ are functions on the shape space $\widehat Q$.

\item[{\bf Step 3.}] A necessary condition for the existence of an invariant measure  is that 
\begin{equation*}
C_{a b}^b+C_{a \alpha}^\alpha =0, \qquad \mbox{for all $a$}.
\end{equation*}

We shall see that the latter condition only holds if two of the semi-axes of the ellipsoid are equal.
\end{enumerate}

A rough explanation of the ideas behind the steps of the  algorithm described above is presented in the appendix.

\section{A homogeneous ellipsoid rolling without slipping on the plane}
\label{S:examples}

Consider the motion of a homogeneous ellipsoid that rolls without slipping on the plane. 
We assume that its semi-axes have lengths $a,b,c>0$.
If two of the semi-axes have equal length, e.g. the ellipsoid is a solid of revolution, then there
exists an invariant measure, see e.g. \cite{BM}.

 The space frame $\{e_1, e_2, e_3 \}$ is chosen so that the rolling takes place on the plane
$z=0$. 
 We consider a body frame $\{E_1, E_2, E_3 \}$, whose origin is located at center $O$ of the ellipsoid and is 
 aligned with the principal axes of symmetry of the body.  We denote by $\bf r$ the vector that connects
 $O$ with the contact point $P$ of the ellipsoid and the plane written in body coordinates, and by
 $\boldsymbol{\gamma}$ the {\em Poisson vector} that is the unit normal vector $e_3$ to the plane written in body coordinates.
 See figure \ref{F:ellipsoid}.

\begin{figure}[ht]
\centering
\includegraphics[width=18cm]{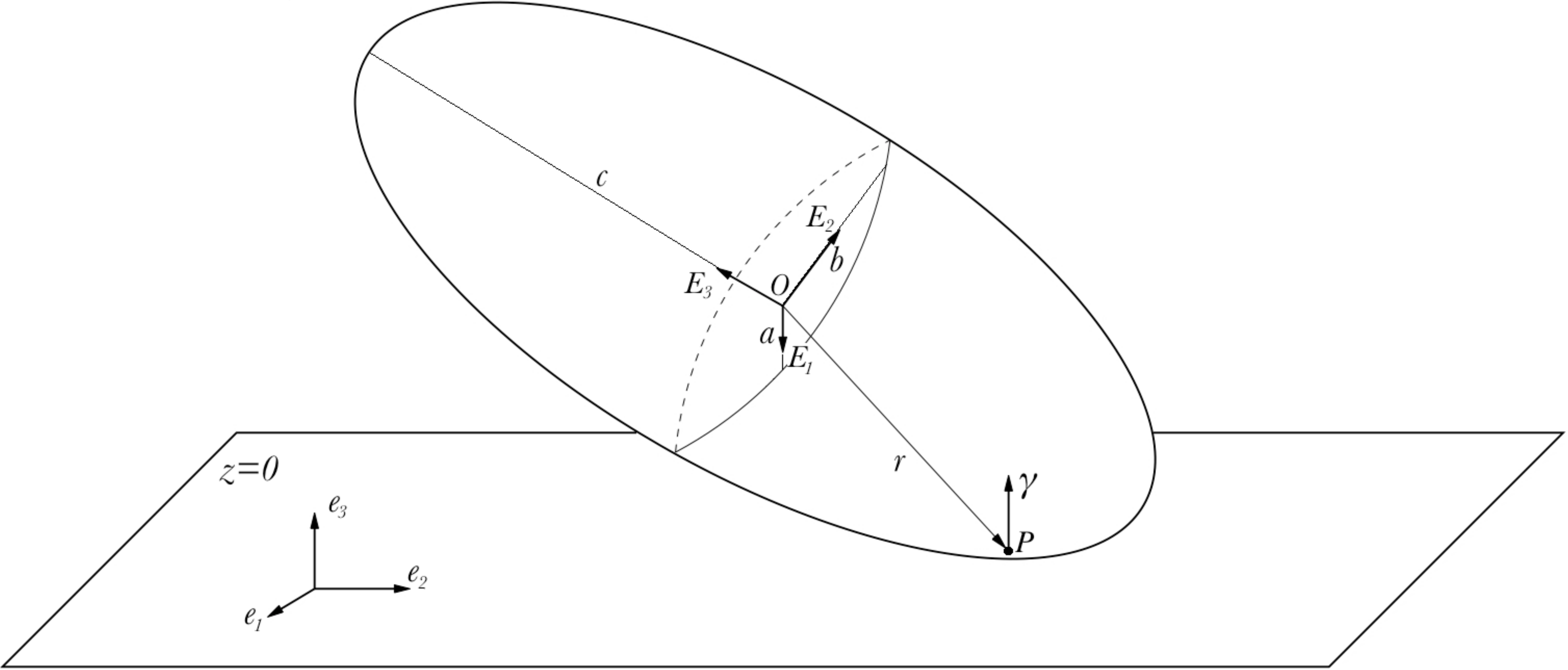}
\caption{\small{Ellipsoid rolling on the plane}\label{F:ellipsoid} }
\end{figure}

The vectors $\bf r$ and  $\boldsymbol{\gamma}$ are related by:
\begin{equation*}
\begin{split}
{\bf r}&=\left (\frac{-a^2\gamma_1}{\sqrt{a^2\gamma_1^2+b^2\gamma_2^2+c^2\gamma_3^2}},\frac{-b^2\gamma_2}{\sqrt{a^2\gamma_1^2+b^2\gamma_2^2+c^2\gamma_3^2}},\frac{-c^2\gamma_3}{\sqrt{a^2\gamma_1^2+b^2\gamma_2^2+c^2\gamma_3^2}} \right )^t , \\
\boldsymbol{\gamma}&=\left (\frac{-b^2c^2r_1}{\sqrt{b^4c^4r_1^2+a^4c^4r_2^2+a^4b^4r_3^2}},\frac{-a^2c^2r_2}{\sqrt{b^4c^4r_1^2+a^4c^4r_2^2+a^4b^4r_3^2}},\frac{-a^2b^2r_3}{\sqrt{b^4c^4r_1^2+a^4c^4r_2^2+a^4b^4r_3^2}} \right ) ^t .
\end{split}
\end{equation*}

Denote by ${\bf x}=(x,y,z)^t$ the spatial coordinates of the center of the ellipsoid.
 A matrix $g\in SO(3)$ specifies the orientation of the ellipsoid by relating the body and the space frame. 
 The Poisson vector $\boldsymbol{\gamma}=g^{-1}e_3$.
 The constraint of rolling without slipping is expressed by the vectorial relation $\dot {\bf x}=-\dot g{\bf r}$.
 This vectorial constraint includes the holonomic constraint $z=-{\bf r}\cdot \boldsymbol{\gamma}$ where
 ``$\cdot$" denotes the euclidean inner product in $\R^3$
 (to see this note that $\dot {\bf r}\cdot \boldsymbol{\gamma}=0$).
 Therefore, the configuration space is $Q=\R^2\times SO(3)$ where $(x,y)$ are coordinates in the $\R^2$
 part.
 We will use Euler angles as local
coordinates for 
$SO(3)$. We use  the {\em $x$-convention}, see e.g. \cite{MaRa} and write a matrix $g\in SO(3)$ 
as 
\begin{equation*}
g=\left(
\begin{array}{ccc}
 \cos \psi \cos \varphi - \cos \theta \sin \varphi \sin \psi & -\sin \psi \cos \varphi - \cos \theta \sin \varphi \cos \psi & \sin \theta \sin \varphi \\
\cos \psi \sin \varphi + \cos \theta \cos \varphi  \sin \psi  & -\sin \psi \sin \varphi + \cos \theta \cos \varphi  \cos \psi  & -\sin \theta \cos \varphi  \\
 \sin \theta \sin \psi   & \sin \theta \cos \psi  & \cos \theta  
\end{array}
\right),
\end{equation*}
where the Euler angles $0<\varphi , \psi <2\pi, \, 0<\theta <\pi$. 
According to this convention, we
obtain $\boldsymbol{\gamma}=(\sin \theta \sin \psi, \sin \theta \cos \psi, \cos \theta)^t$.

The holonomic constraint, coming from the third  component
of the  relation  $\dot {\bf x}=-\dot g{\bf r}$, is explicitly given by
\begin{equation*}
z=z(\theta,\psi)=\sqrt{a^2\sin^2\theta\sin^2\psi +b^2\sin^2\theta\cos^2\psi + c^2\cos^2\theta}.
\end{equation*}

The nonholonomic constraints of rolling without slipping, coming from the first two components
of the  relation  $\dot {\bf x}=-\dot g{\bf r}$, are explicitly given by
\begin{equation*}
\label{E:Constraints-rolling-Chap-top}
\begin{split}
\dot x &= A(\varphi, \theta, \psi) \, \dot \varphi 
+z(\theta, \psi)\sin\varphi \, \dot \theta  + E(\varphi, \theta, \psi) \,\dot \psi,\\
\dot y &=B(\varphi, \theta, \psi) \, \, \dot \varphi 
-z(\theta, \psi)\cos\varphi \, \dot \theta + F(\varphi, \theta, \psi) \, \dot \psi, 
\end{split}
\end{equation*}
where
\begin{equation*}
\begin{split}
A(\varphi, \theta, \psi) \,&=\frac{\sin\theta}{z(\theta, \psi)}
\left (a^2(-\sin \psi \cos \psi\sin \varphi-\cos \theta \cos \varphi \sin^2\psi  )\right . \\ & \qquad \qquad \qquad 
\qquad \left .+b^2(  \cos \psi
\sin \psi \sin \varphi - \cos ^2 \psi \cos \theta \cos \varphi )+c^2 \cos \varphi \cos \theta \right ), \\
B(\varphi, \theta, \psi) \,&= \frac{\sin\theta}{z(\theta, \psi)}
\left (a^2(\sin \psi \cos \psi\cos \varphi-\cos \theta \sin \varphi \sin^2\psi  )\right . \\ & \qquad \qquad \qquad 
\qquad \left . +b^2( - \cos \psi
\sin \psi \cos \varphi - \cos ^2 \psi \cos \theta \sin \varphi )+c^2 \sin \varphi \cos \theta \right ),
\end{split}
\end{equation*}
\begin{equation*}
\begin{split}
E(\varphi, \theta, \psi) \,&=\frac{\sin \theta}{z(\theta,\psi)} \left (a^2(-\sin^2 \psi \cos  \varphi-\sin \psi \cos \theta \sin \varphi \cos \psi)\right . \\ & \qquad \qquad \qquad 
\qquad \left . +b^2( - \cos^2 \psi
\cos \varphi + \cos \psi \cos \theta \sin \varphi  \sin \psi )\right), \\
F(\varphi, \theta, \psi) \,&=\frac{\sin \theta}{z(\theta,\psi)} \left (a^2(-\sin^2 \psi \sin  \varphi+\sin \psi \cos \theta \cos \varphi \cos \psi)\right . \\ & \qquad \qquad \qquad 
\qquad \left . +b^2( - \cos^2 \psi
\sin \varphi - \cos \psi \cos \theta \cos \varphi  \sin \psi )\right).
\end{split}
\end{equation*}

The kinetic energy of the ellipsoid is given by
\begin{equation}
\label{E:KinEnergy_ChapTop}
\mathcal{K}=\frac{1}{2}\langle \I \boldsymbol{\Omega} , \boldsymbol{\Omega} \rangle +\frac{m}{2} || \dot {\bf x}||^2,
\end{equation}
where  $m$ is the total mass of the ellipsoid,
\begin{equation*}
\I=\frac{m}{5}\left ( \begin{array}{ccc} b^2+c^2 & 0 & 0 \\ 0 & a^2+c^2 & 0 \\ 0 & 0 & a^2+b^2 \end{array} \right )
\end{equation*}
 is the inertia tensor of the ellipsoid with respect to $O$ and with our choice of body axes.
In order to think of $\mathcal{K}$ as a Riemannian metric $\mathcal{G}$ in $Q=SO(3)\times \R^2$ it is understood that one needs to put $\dot z = \frac{\partial z}{\partial \theta} \dot \theta+  \frac{\partial z}{\partial \psi} \dot \psi$ in the expression for the kinetic energy \eqref{E:KinEnergy_ChapTop}.
 The vector $ \boldsymbol{\Omega}$ is the angular velocity of the sphere written in body coordinates and
 in terms of Euler angles is given by
 \begin{equation*}
\label{E:Omega-omega}
  \boldsymbol{\Omega}=\left ( \dot \theta \cos \psi +\dot \varphi \sin \psi \sin \theta \, , \, -\dot \theta \sin \psi + \dot \varphi \cos \psi \sin \theta  \, , \,  \dot 
\varphi\cos \theta + \dot \psi  \right  )^t.
\end{equation*}

\medskip{{\bf Symmetries}}

 There is a freedom in the choice of origin and orientation of the space axes $\{ e_1, e_2 \}$. This corresponds to a symmetry of the system defined by a left action of the Euclidean group $SE(2)$ on $Q$. 
 Let 
 \begin{equation*}
h=\left ( \begin{array}{ccc} \cos \vartheta & -\sin \vartheta& v \\  \sin \vartheta & \cos \vartheta& w \\ 0 & 0 & 1 \end{array} \right )
\end{equation*}
denote a generic element on $SE(2)$. The action of $h$ on a point $q\in Q$ with local coordinates
$(\varphi, \theta, \psi, x,y)$ is given by
\begin{equation*}
h\cdot  q = (\varphi + \vartheta ,  \theta, \psi, x \cos\vartheta  - y\sin \vartheta +v , x \sin \vartheta  + y\cos \vartheta +w).
\end{equation*}
One can check that both the constraints and the kinetic energy are invariant under the lift of the action
to $TQ$. The action of $SE(2)$ on $Q$ is free and proper and the shape space $\widehat Q =Q/G =S^2$.
In our local coordinates the orbit projection $p:Q\to S^2$ is given by
\begin{equation*}
p(\varphi, \psi, \theta, x,y) = ( \theta, \psi),
\end{equation*} 
where $( \theta, \psi)$ are spherical coordinates on the unit sphere $\gamma_1^2+\gamma_2^2+\gamma_3^2=1$, defined by
\begin{equation*}
\gamma_1=\sin \theta \sin \psi, \qquad \gamma_2= \sin \theta \cos \psi, \qquad  \gamma_3= \cos \theta.
\end{equation*}

The vertical subbundle $\mathcal{V}p$ is spanned by 
\begin{equation*}
\mathcal{V}p=\mbox{span} \left \{ \frac{\partial}{\partial \varphi}\,  , \,    \frac{\partial}{\partial x} \,  , \,  
 \frac{\partial}{\partial y} \right \}.
\end{equation*}
On the other hand, the constraint distribution $D$ is spanned by the $SE(2)$- invariant vector fields
\begin{equation*}
\begin{split}
Z_{a=1}&= \frac{\partial}{\partial \varphi} + A\frac{\partial}{\partial x} +B\frac{\partial}{\partial y} \,  , \\  X_{\alpha=1}&=\frac{\partial}{\partial \theta} +z(\theta, \psi)\sin \varphi 
 \frac{\partial}{\partial x} - z(\theta, \psi)\cos \varphi 
 \frac{\partial}{\partial y}\,  , \\ X_{\alpha=2}&=\frac{\partial}{\partial \psi}  +E \frac{\partial}{\partial x}+ F \frac{\partial}{\partial y}.
 \end{split}
\end{equation*}
It is then clear that the intersection $\mathcal{V}^Dp=D \cap \mathcal{V}p$ has constant rank 1 and is spanned by $Z_{a=1}$. 

The following vector fields, together with $Z_{a=1}$ satisfy the requirements of step 1 of the algorithm:
\begin{equation*}
Y_{\alpha} := X_\alpha - \frac{\mathcal{G}(Z_{a=1},X_\alpha)}{\mathcal{G}(Z_{a=1},Z_{a=1})} \,
Z_{a=1} \, \qquad \alpha=1, 2.
\end{equation*}

Since  the sub-index $a$ only takes the value 1, and the sub-indices $\alpha, \beta$ only take the  values $1,2$ the 
condition in step 3 of the algorithm simplifies to 
\begin{equation*}
C_{a=1, \alpha=1}^{\alpha=1}+C_{a=1, \alpha=2}^{\alpha=2}=0.
\end{equation*}

We will now study the above condition. We start by computing the standard commutators:
\begin{equation}
\label{E:Commutators}
\begin{split}
[Z_{a=1}, Y_{\alpha=1}] &= [Z_{a=1}, X_{\alpha=1}]+ \lambda_1Z_{a=1} \, , \\
[Z_{a=1}, Y_{\alpha=2}] &= [Z_{a=1}, X_{\alpha=2}]+ \lambda_2 Z_{a=1},
\end{split}
\end{equation}
where  
\begin{equation*}
\lambda_1=Z_{a=1}\left ( - \frac{\mathcal{G}(Z_{a=1},X_{\alpha=1})}{\mathcal{G}(Z_{a=1},Z_{a=1})} \right ), \qquad \lambda_2=Z_{a=2}\left ( - \frac{\mathcal{G}(Z_{a=1},X_{\alpha=1})}{\mathcal{G}(Z_{a=1},Z_{a=1})} \right ).
\end{equation*}
are  functions of $(\varphi,\theta, \psi)$. We should now
compute the $\mathcal{G}$-orthogonal projection of the commutators in equation \eqref{E:Commutators}  onto $D$ and express them
as a linear combination of $Z_{a=1}, Y_{\alpha=1}, Y_{\alpha=2}$ to determine the
coefficients $C_{a=1, \alpha}^I$. In fact,  looking ahead at step 3 of the algorithm,
we are interested in computing $C_{a=1, \alpha}^\alpha$ for $\alpha=1,2$.
A simple linear algebra argument shows that $C_{a=1, \alpha=1}^{\alpha=1}$
 coincides with the component of $X_{\alpha=1}$ when the $\mathcal{G}$-orthogonal projection
 of 
 \begin{equation*}
[Z_{a=1}, X_{\alpha=1}]=\left ( z(\theta, \psi) \cos \varphi -\frac{\partial A}{\partial \theta} \right ) \frac{\partial}{\partial x} + \left ( z(\theta, \psi) \sin \varphi -\frac{\partial B}{\partial \theta} \right )  \frac{\partial}
{\partial y}
\end{equation*}
onto $D$ is expressed in terms of the basis $Z_{a=1}, X_{\alpha=1}, X_{\alpha=2}$.
The same idea can be used to compute $C_{a=1, \alpha=2}^{\alpha=2}$. Using these observations
 and with the aid of MAPLE\texttrademark\, we obtain:
 \begin{equation}
 \label{E:trace-Chap-Top}
C_{a=1, \alpha=1}^{\alpha=1}+C_{a=1, \alpha=2}^{\alpha=2}=\frac{3m^3 \sin^4 \theta \cos \theta\sin \psi \cos \psi }{50\, \mbox{det}
(T)z(\theta,\psi)^4} (a^2-b^2)(b^2-c^2)(c^2-a^2) G(\theta, \psi),
\end{equation}
where $T$ is the (positive definite) matrix 
\begin{equation*}
T=\left ( \begin{array}{ccc} 
\mathcal{G}( Z_{a=1},  Z_{a=1}) & \mathcal{G}( Z_{a=1},  X_{\alpha=1}) &  \mathcal{G}( Z_{a=1},  X_{\alpha=2}) \\
 \mathcal{G}( Z_{a=1},  X_{\alpha=1}) &\mathcal{G}( X_{\alpha=1},  X_{\alpha=1})&\mathcal{G}( X_{\alpha=1},  X_{\alpha=2})
 \\
  \mathcal{G}( Z_{a=1},  X_{\alpha=2}) &\mathcal{G}( X_{\alpha=1},  X_{\alpha=2})&\mathcal{G}( X_{\alpha=2},  X_{\alpha=2})
 \end{array} \right ),
\end{equation*}
 and the  function $G$ is given by:
 \begin{equation*}
\begin{split}
G( \theta, \psi)&=(2a^2b^2+3a^2c^2+3b^2c^2) +(b^2-a^2)c^2 (1-\cos(2\theta)) \cos(2\psi) \\
& \qquad \qquad + \left (-2a^2b^2+a^2c^2+b^2c^2 \right ) \cos(2\theta).
\end{split}
\end{equation*}

 The necessary condition for the existence of an invariant measure, coming from step 3
 of the algorithm, is that the expression \eqref{E:trace-Chap-Top}
 vanishes identically for all $(\theta, \psi)$ in the chart, that is, for all $0<\theta < \pi$, $0<\psi<2\pi$. It
 is easily seen that $G(  \theta, \psi)$ is not identically zero for any allowed values 
 of the parameters $a, b$ and $c$. Hence, the quantity \eqref{E:trace-Chap-Top} can only be
 identically zero if any two out of the three semi-axis' lengths $a, b$ and $c$ are equal.
 As mentioned before, in this case the ellipsoid is a solid of revolution and it is known  \cite{Yar} (see also \cite{BM}) that an invariant measure exists.
It is shown in \cite{BM} that the reduced equations of motion, for an arbitrary shape of the 
ellipsoid, can be presented in the vectorial form
 \begin{equation*}
 \label{E:Motion-Chap-Top}
\dot {\bf K} = {\bf K} \times \boldsymbol{\Omega} + m\dot {\bf r}  \times ( \boldsymbol{\Omega} \times{\bf r}), \qquad \dot{\boldsymbol{\gamma}}=\boldsymbol{\gamma}\times \boldsymbol{\Omega},
\end{equation*}
and in the case when $a=b$ possess the invariant measure:
\begin{equation*}
\frac{d\boldsymbol{\gamma} \wedge d{\bf K} }{\sqrt{\frac{2}{25}m^2a^2(a^2+c^2)+m\langle {\bf r}, \I  {\bf r} \rangle}} = \left (\frac{m}{5}(a^2+c^2)+m|| {\bf r}||^2 \right )\sqrt{\frac{2}{25}m^2a^2(a^2+c^2)+m\langle{\bf r}, \I  {\bf r} \rangle} \, d\boldsymbol{\gamma} \wedge d\boldsymbol{ \Omega}\, .
\end{equation*}
In the above formulae  ${\bf K}$ is the angular momentum of the ellipsoid with respect
to the contact point, also written with respect to the body frame. Explicitly we have:
\begin{equation*}
 \qquad {\bf K}=\I \boldsymbol{ \Omega} + m {\bf r} \times (\boldsymbol{ \Omega}\times  {\bf r}).
\end{equation*}

 Therefore we have:
\begin{theorem}
\label{T:Chaplygin-top}
The reduced equations for a homogeneous ellipsoid that rolls without slipping on the plane possess an invariant measure if and only if at least two of its semi-axes are equal.
\end{theorem}
We stress that the above conditions were known to be sufficient but we have shown that they are also
necessary. 

\begin{remark}
\label{R:Basicisenough}
The content of Theorem \ref{T:Chaplygin-top} applies to general smooth measures, including
those with velocity dependent densities. This is a consequence of Theorems 3.6 and 3.8 in
 \cite{FeGaMa}.
 \end{remark}

\renewcommand{\thesection}{\sc{Appendix}}

\section{Some comments on the algorithm to study the existence of invariant measures}
\label{A:Volume-forms}

\renewcommand{\thesection}{A}

Our method relies on the form of the reduced equations of motion when we work in quasi-velocities $ v^a, \, v^\alpha$ defined with respect to
the basis of vector fields $\{Z_a, Y_\alpha \}$. The details are given in \cite{FeGaMa}, but for completeness we present a rough exposition in this
appendix. We keep the same notation used in section \ref{S:algorithm}. In particular, we keep using the same conventions on the use
of the sub-indices $a, b, e$, $\alpha, \beta, \gamma$,  and $I, J, K$.

The kinetic energy $\mathcal{K}$ of the system, being $G$-invariant, can be expressed in terms of the quasi-velocities $ v^a, \, v^\alpha$ as
\begin{equation*}
\mathcal{K}=\frac{1}{2} \left (K_{ab}(\hat q^i) v^av^b +K_{\alpha \beta}(\hat q^i)v^\alpha v^\beta \right ),
\end{equation*}
where $\hat q^i$ are coordinates of the shape space $\widehat Q$. In the above expression $K_{ab}(\hat q^i)$ and $K_{\alpha \beta}(\hat q^i)$
are everywhere positive definite matrices. Notice that there are no crossed terms $v^\alpha v^a$ in the kinetic energy since the vector fields
$\{Z_a\}$ and  $\{Y_\alpha \}$ were chosen to be $\mathcal{G}$-orthogonal.

Define the generalized momentum variables $p_a, \, p_\alpha$ and the Hamiltonian $H$ of the system as usual,
\begin{equation*}
p_a=\frac{\partial \mathcal{K}}{\partial v^a}, \qquad p_\alpha =\frac{\partial \mathcal{K}}{\partial v^\alpha},  \qquad
H=\frac{1}{2} \left (K^{ab}(\hat q^i) p_ap_b +K^{\alpha \beta}(\hat q^i)p_\alpha p_\beta \right ),
\end{equation*}
 where the matrices $K^{ab}(\hat q^i)$ and $K^{\alpha \beta}(\hat q^i)$ are  inverses of $K_{ab}(\hat q^i)$ and $K_{\alpha \beta}(\hat q^i)$ respectively.
With respect to the coordinates $\hat q^i$ and the generalized momenta $p_a, \, p_\alpha$, the reduced equations of motion take
the form:
\begin{equation}
\begin{split}
\label{E:Motion}
\frac{d\hat q^i}{dt}& = \rho^i_\beta \frac{\partial H}{\partial p_\beta}, \\ \frac{d p_\alpha}{dt}& =- \rho^i_\alpha \frac{\partial H}{\partial \hat q^i}
-C_{\alpha \beta}^\gamma p_\gamma \frac{\partial H}{\partial p_\beta}-C_{\alpha \beta}^a p_a \frac{\partial H}{\partial p_\beta}-C_{\alpha b}^\beta p_\beta \frac{\partial H}{\partial p_b}-C_{\alpha b}^ap_a \frac{\partial H}{\partial p_b} ,\\
 \frac{dp_a}{dt} &=
-C_{a \beta}^\alpha p_\alpha \frac{\partial H}{\partial p_\beta}-C_{a \beta}^b p_b \frac{\partial H}{\partial p_\beta}-C_{ab}^\alpha p_\alpha \frac{\partial H}{\partial p_b}-C_{ a b}^e p_e \frac{\partial H}{\partial p_b}.
\end{split}
\end{equation}
In the above equations the coefficients $C_{\alpha \beta}^I$ are defined by the relation
\begin{equation*}
\mathcal{P}([Y_\alpha, Y_\beta])=C_{\alpha \beta}^aZ_a + C_{\alpha \beta}^\gamma Y_\gamma ,
\end{equation*}
and the coefficients $\rho_\alpha^i$ satisfy
\begin{equation*}
Tp(Y_\alpha)=\rho_\alpha^i\partial_{\hat q^i},
\end{equation*}
where we recall that $p:Q\to \widehat Q$ is the orbit projection.

Now the key point is that, since the coefficients $\rho_\alpha^i,\, C_{\alpha I}^J$ and $C_{aI}^J$ only depend in $\hat q^i$, the second and third of equations
\eqref{E:Motion} are homogeneous quadratic in the momentum variables $p_a, \, p_\alpha$. As a consequence (see \cite{FeGaMa} and Remark \ref{R:Basicisenough}), it suffices to search for
basic measures of the form
\begin{equation*}
e^{\sigma(\hat q^i)}\, d\hat q^i \, dp_\alpha \, dp_a,
\end{equation*}
where the smooth function $\sigma$ does not depend on $p_a, \, p_\alpha$. Taking the divergence of the vector field defined by 
equations \eqref{E:Motion}  with respect to the above measure and equating to zero yields the following equation for $\sigma$:
\begin{equation*}
\left ( \rho_\beta^i\frac{\partial \sigma}{\partial q^i} +\frac{\partial  \rho_\beta^i}{\partial q^i}-C_{\alpha \beta}^\alpha -C_{a \beta}^a\right ) \frac{\partial H}{\partial p_\beta} +\left (-C_{\alpha b}^\alpha - C_{ab}^a\right ) \frac{\partial H}{\partial p_b}=0,
\end{equation*}
where we have used the equality of mixed partial derivatives and the skew-symmetry of the coefficients $C_{IJ}^K$ with respect to the
lower indices. Differentiating the above equation with respect to $p_e$ and using the block diagonal form of the Hamiltonian $H$ gives
\begin{equation*}
K^{eb}(\hat q^i)\left (C_{\alpha b}^\alpha + C_{ab}^a\right )=0.
\end{equation*}
The condition appearing in step 3 of the algorithm now follows by using the invertibility of the matrix $K^{eb}(\hat q^i)$.

\end{document}